\documentclass[11pt]{article}
\usepackage{moriond,epsfig}

\def\be{\begin{equation}}
\def\ee{\end{equation}}
\def\bea{\begin{eqnarray}}
\def\eea{\end{eqnarray}}

\begin{document}
\vspace*{2.8cm}
\title{$B_s$ MIXING, $\Delta\Gamma_s$ AND {\sf CP}~VIOLATION \\
       AT THE TEVATRON}

\author{Gian Piero DI GIOVANNI, \\ for the CDF and D\O~Collaborations \\\,}
 
\address{LPNHE, Laboratoire de Physique Nucl\'eaire et des Hautes Energies,\\ 
         4 Place Jussieu, Tour 33 RdC, 75005,\\ 
         Paris, France}

\maketitle\abstracts{
We discuss the results from the Tevatron experiments on mixing and  {\sf CP}
violation in the $B_s^0-\bar{B}_s^0$ system, with particular emphasis to the 
updated measurements of the decay-width difference $\Delta\Gamma_s$ and the 
first measurement of the {\sf CP}-violating phase $\beta_s$ using flavor tagging
information. We also briefly review the charge asymmetry measurements in 
semileptonic $B_s^0$ decays and in $B^\pm \to J/\psi K^\pm$ decays.}

%===============================================================================
\section{Introduction}
\label{sec:intro}
%===============================================================================
The Tevatron is a $p\bar{p}$ collider operating at the Fermi National 
Accelerator Laboratory. The protons and anti-protons collide at a center-of-mass
energy of $\sqrt{s}=1.96 \mbox{ TeV}$ in two interaction points, where the CDF 
II and D\O~detectors are located. The two experiments have collected an 
integrated luminosity of $3~\mathrm{fb}^{-1}$ and the measurements presented 
here span from $1.0~\mathrm{fb}^{-1}$ to $2.8~\mathrm{fb}^{-1}$.
The physics of the $b$ quark is a very active research area to challenge the 
Standard Model predictions. Precise measurements in $B^0$ and $B^+$ meson 
decays, performed at the $B$ factories, improved the understanding of flavor 
dynamics and proved the Standard Model description very successful. On the other
hand, a comparable experimental knowledge of $B_s^0$ decays has been lacking. 
The $B_s^0$ oscillation observation at CDF~\cite{Abulencia:2006ze} 
strongly constrained the magnitude of New Physics contributions in the $B_s^0$
mixing, while its phase, responsible for {\sf CP} violating effects, is not 
precisely determined yet. The $B_s^0$ sector offers a large variety of 
interesting processes in which large {\sf CP} violation effects are still 
allowed by the current experimental constraints, but are negligible small in the
Standard Model.
Thus, the Tevatron collider, providing a simultaneous access to large samples of
strange and non-strange $b$-mesons necessary for precision measurements, offers
a great opportunity to study the $B_s^0$ flavor sector, before the start-up of 
CERN Large Hadronic Collider (LHC).

%===============================================================================
\section{Phenomenology of the $B_s^0$ System}
\label{sec:bspheno}
%===============================================================================
Flavor oscillation, or mixing, is a very well established phenomenon in 
particle physics.
In the Standard Model the mass and the flavor eigenstates of neutral $B$ mesons
differ. This give rise to particle-antiparticle oscillations, which proceed
through forth-order flavor changing weak interactions, whose phenomenology 
depends on the Cabibbo-Kobayashi-Maskawa (CKM) quark mixing matrix.
The rate at which the neutral $B-\bar{B}$ transitions occur is governed by the 
mass difference, $\Delta m$ of the two mass eigenstates, $B^L$ and
$B^H$, where the superscripts $L$ and $H$ stay for ``light'' and ``heavy''. 
The phenomenology of mixing in $B_s^0$ and $\bar{B}_s^0$ mesons is, then,
characterized by the mass difference of the two mass eigenstates, $\Delta m_s$,
as well as by the decay width-difference
$\Delta\Gamma_s \equiv \Gamma_s^L - \Gamma_s^H = 1/\tau_{B_s^L}-1/\tau_{B_s^H}$.
The latter depends on the {\sf CP} violating phase defined as 
$\phi_s=\arg(-M_{12}/\Gamma_{12})$, through the relationship 
$\Delta\Gamma_s=2|\Gamma_{12}|\times\cos(\phi_s)$. $M_{12}$ and $\Gamma_{12}$ 
are the off-diagonal elements of the $B_s^0-\bar{B}_s^0$ decay matrix from the 
Schr\"{o}edinger equation describing the time evolution of $B_s^0$ mesons
\cite{Dighe:1998vk}$^,$~\cite{Dunietz:2000cr}. While the Standard Model 
expectations are small~\cite{Lenz:2006hd}, $\phi_s=4 \times 10^{-3}$, New 
Physics could significantly modify the observed phase value contributing with 
additional processes, $\phi_s=\phi_s^{SM}+\phi_s^{NP}$. The same phase would 
alter the observed phase between the mixing and the $b \to c\bar{c}s$ 
transitions,
$2\beta_s=2\beta_s^{SM}-\phi_s^{NP}$, in which the Standard Model contribution 
is defined as $-2\beta_s^{SM}=-2\arg(-\frac{V_{ts}V_{tb}^*}{V_{cs}V_{cb}^*})
\approx \mathcal{O}(0.04)$, where $V_{ij}$ are the elements of the CKM matrix.
Since both $\phi_s^{SM}$ and $\beta_s^{SM}$ are tiny with respect to 
the current experimental resolution, we can approximate $\phi_s=-2\beta_s$.
A measurement of sizable value of $2\beta_s$ ($\phi_s$) would be a clear 
indication of New Physics~\cite{Dighe:1998vk}$^,$~\cite{Dunietz:2000cr}.

%===============================================================================
\section{$B_s^0$ Mixing}
\label{sec:bsmix}
%===============================================================================
While $\Delta m_d$ was precisely determined at the $B$ factories
~\cite{Aubert:2001te}$^,$~\cite{Tomura:2002qs}, the $B_s^0$ mixing frequency has
been first measured at CDF experiment~\cite{Abulencia:2006ze}.
The $B_s^0-\bar{B}_s^0$ oscillation observation was achieved through a 
combination of several data-sets of $1~\mathrm{fb}^{-1}$, in integrated 
luminosity, which results in:

\begin{equation}
\Delta m_s = 17.77 \pm 0.10 \mbox{ (stat.)} \pm 0.07 \mbox{  (syst.)}
             ~\mathrm{ps}^{-1},
\end{equation}

with a significance greater than $5$ standard deviations. 
Two independent types of flavor tags are used to identify the $B_s^0$ flavor at
production: the Opposite Side Tagger (OST) and the Same Side Kaon Tagger (SSKT).
The performance of flavor taggers are quantified by the efficiency $\epsilon$ 
and the dilution $\mathcal{D}$, defined as the probability to correctly tag a 
candidate. The tagging effectiveness, $\epsilon \mathcal{D}^2$ of the OST is
$1.8\%$. The SSKT has $\epsilon \mathcal{D}^2 = 3.5\%$ (hadronic) and 
$4.8\%$ (semileptonic) and thus contributes most to the sensitivity of the CDF
analysis.
The accurate measurement of the $B_s^0-\bar{B}_s^0$ mixing frequency offers a 
powerful constraint to the ratio $|V_{ts}|^2/|V_{td}|^2$ of CKM matrix elements:

\begin{equation}\label{eqn:ratioVtsVtd}
\frac{|V_{ts}|^2}{|V_{td}|^2} = 0.2060 \pm 0.0007 \mbox{ (stat.)}
^{+0.0081}_{-0.0060} \mbox{ (theory) }.
\end{equation}

D\O~recently reported a measurement of the $B_s^0$ oscillation frequency
~\cite{bsmixD0} using a large sample of semileptonic $B_s^0$ decays and their 
first hadronic mode, $B_s^0 \to D_s^- [\to \phi(\to K^+ K^-)~\pi^-]~\pi^+$. 
D\O~combines the tagging algorithms using a likelihood-ratio method, obtaining 
a total effective tagging power $\epsilon \mathcal{D}^2 =(4.49 \pm 0.88)\%$.
With a data-set of approximately $2.4~\mathrm{fb}^{-1}$, they obtains: 

\begin{equation}
\Delta m_s = 18.56 \pm 0.87 \mbox{ (stat.)} \mathrm{ps}^{-1}.
\end{equation}

The result statistically exceeds the $3\sigma$ significance and it is 
compatible with the CDF measurement. The $\Delta m_s$ is well consistent with 
the Standard Model unitarity hypothesis for the CKM matrix. 
 
%===============================================================================
\section{Phase of the Mixing Amplitude and Decay-Width Difference in the $B_s^0$
System}
\label{sec:DGs_BETS}
\begin{figure}
\begin{center}
\epsfig{figure=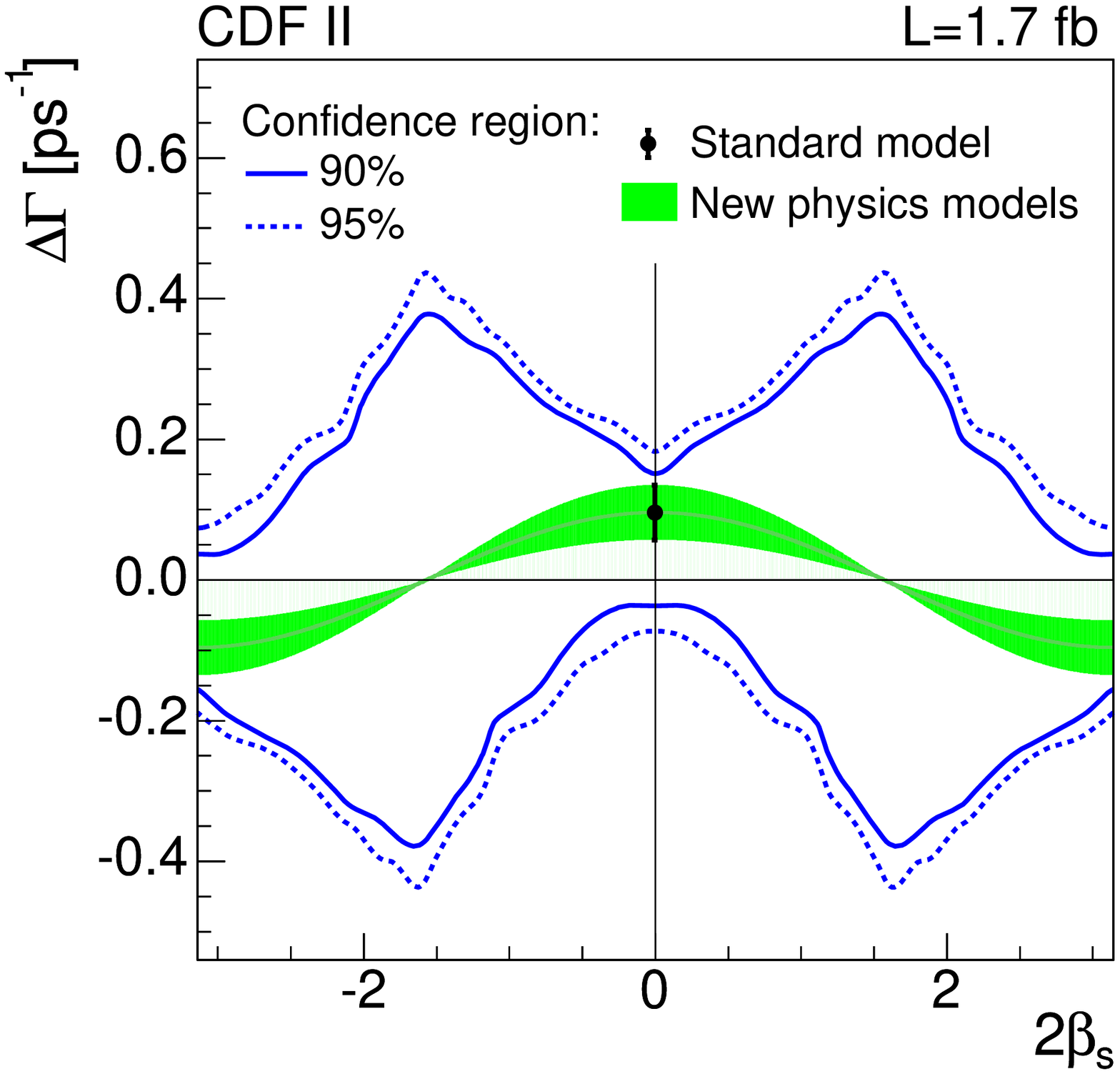,height=1.9in, width=2.2in}
\epsfig{figure=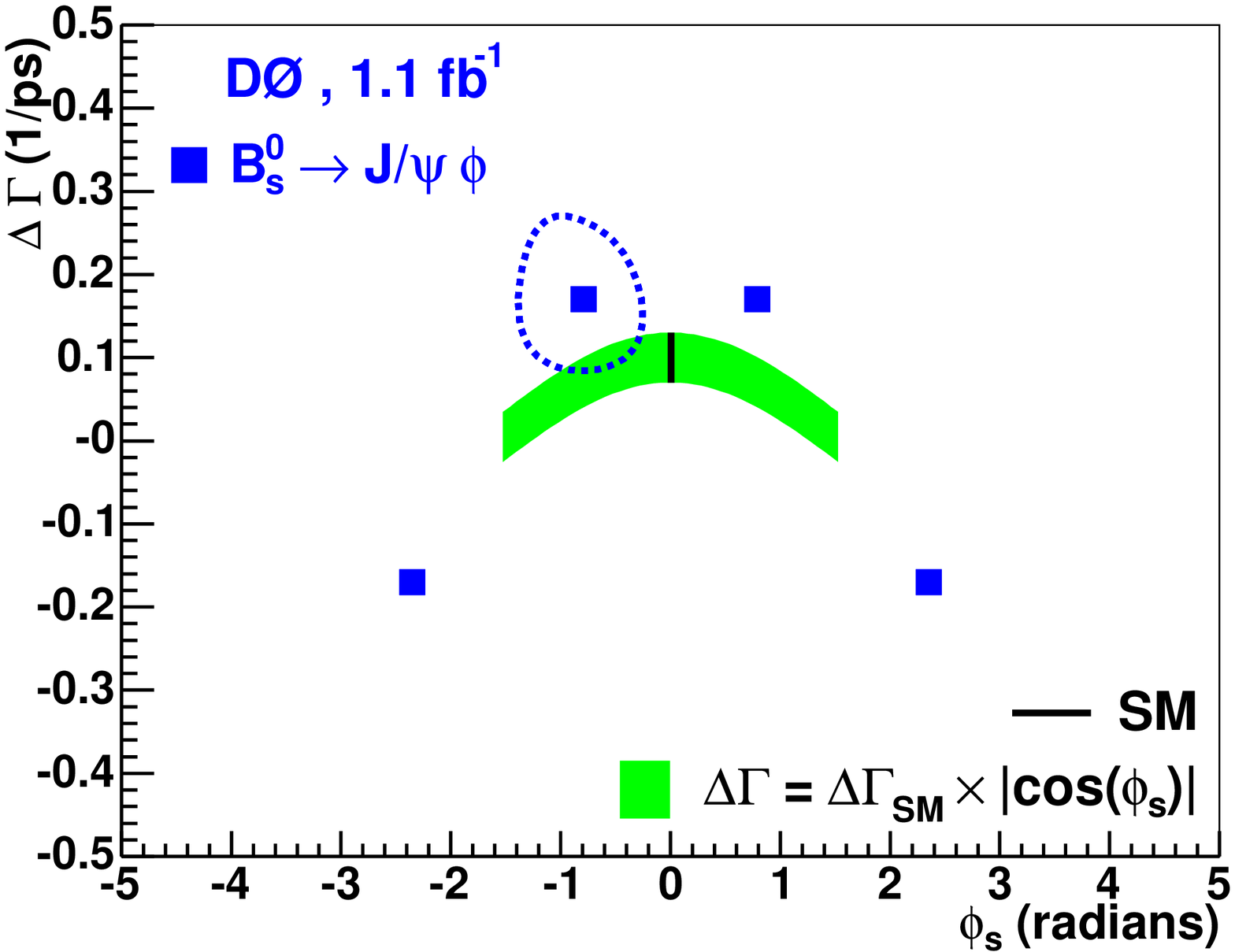         ,height=1.8in, width=2.2in}
\caption{Left: CDF $90\%$ (solid) and $95\%$ (dashed) confidence level contour 
  in the plane $2\beta_s-\Delta\Gamma_s$, compared with the SM prediction and 
  the region allowed in New Physics model given by 
  $\Delta\Gamma_s=2|\Gamma_{12}|cos(2\beta_s)$, with 
  $\Gamma_{12}=0.048 \pm 0.018~\mathrm{ps}^{−1}$.
  Right: D\O~point estimate in the plane $\Delta\Gamma_s-\phi_s$. The $39\%$ CL
  contour (error ellipse) is additionally drawn in the plane 
  $\phi_s-\Delta\Gamma_s$. It is also shown the band representing the relation 
  $\Delta\Gamma_s=\Delta\Gamma_{SM} \times cos(\phi_s)$, with 
  $\Delta\Gamma_{SM} = 0.10 \pm 0.03 ~\mathrm{ps}^{−1}$. In D\O~nomenclature 
  $\phi_s=-2\beta_s$. The $4$-fold ambiguity is discussed in the text.
  \hspace{15.cm}
  \label{fig:untagged}}
\end{center}
\end{figure}

%===============================================================================
We present the time-dependent angular analyses of 
$B_s^0 \to J/\psi(\to \mu^+ \mu^-)~\phi(\to K^+ K^-)$  decay 
mode performed at the Tevatron experiments. The decay $B_s^0 \to J/\psi \phi$  
proceeds through the $b \to c\bar{c}s$ transition and gives rise to both 
{\sf CP}-even and {\sf CP}-odd final states. Through the angular distributions 
of the $J/\psi$ and $\phi$ mesons, it is possible to statistically separate the 
two final states with different {\sf CP} eigenvalues, thus allowing to 
determine the phase $\beta_s$ and to separate lifetimes for the mass 
eigenstates, so to measure the decay-width difference $\Delta\Gamma_s$. 
After the D\O~analysis~\cite{Abazov:2007tx} of untagged 
$B_s^0 \to J/\psi \phi$ decay sample of $1.1~\mathrm{fb}^{-1}$, and reported at 
Moriond 2007, the CDF Collaboration presents a similar analysis with a sample 
of $1.7~\mathrm{fb}^{-1}$ in integrated luminosity~\cite{2007gf}. 
CDF measures 
$\Delta\Gamma_s=0.076^{+0.059}_{-0.063} \mbox{ (stat.)} \pm 0.006 
                                        \mbox{ (syst.)}~\mathrm{ps}^{-1}$, 
$c\tau_s = 456 \pm 13 \mbox{ (stat.)} \pm 7 \mbox{ (syst.)}~\mu m$, 
assuming {\sf CP} conservation ($\beta_s=0$) results.
To date, this is one of the most precise $B_s^0$ lifetime measurements 
and it is in excellent agreement with both the D\O~results and the theoretical 
expectations predicting $\tau_s = \tau_d \pm \mathcal{O}(1\%)$.
Allowing {\sf CP} violation, a bias and non-Gaussian fit estimates are 
observed in pseudo-experiments for statistics similar to the present data-sets.
The observed bias originates from the
loss of degree of freedom of the likelihood for certain values of the parameters
of interest and does not permit a point estimation of $\Delta\Gamma_s$ and 
$\beta_s$ . 
Thus, CDF provides confidence level regions in the $2\beta_s-\Delta\Gamma_s$ 
plane using the likelihood ratio ordering of Feldman and Cousins~\cite{FC}. For 
the Standard Model expectation 
($\Delta\Gamma_s \approx 0.096~\mathrm{ps}^{-1} \mbox{ and } 
2\beta_s = 0.04~\mathrm{rad}$~\cite{Lenz:2006hd}), the probability to get equal 
or greater likelihood ratio than the one observed in data is $22\%$, which 
corresponds to $1.2$ Gaussian standard deviations.
Figure~\ref{fig:untagged} shows the CDF and the D\O~results in the 
$2\beta_s-\Delta\Gamma_s$ plane.
Furthermore, the CDF Collaboration performed an angular analysis on the 
$B^0 \to J/\psi(\to \mu^+ \mu^-) K^{*0}(\to K^+ \pi^-)$ decay mode for the 
measurement of the transversity amplitudes and strong phases. Such an analysis 
plays a key role in the 
validation of the entire framework used for the $B_s^0 \to J/\psi \phi$ angular 
analysis. The results obtained for the transverse linear polarization amplitudes
at $t=0$, $A_{\parallel}$ and $A_{\perp}$, corresponding to {\sf CP} even and 
{\sf CP} odd final states respectively, as well as the strong phases 
$\delta_{\parallel} = \arg(A_{\parallel}^* A_0)$ and  
$\delta_{\perp}     = \arg(A_{\perp}^* A_0)$, are  
$|A_{\parallel} |^2 = 0.569 \pm 0.009 \mbox{ (stat.)} \pm 0.009\mbox{ (syst.)}$,
$|A_{\perp} |^2 = 0.211 \pm 0.012 \mbox{ (stat.)} \pm 0.006\mbox{ (syst.)}$,  
$\delta_{\parallel} = -2.96 \pm 0.08 \mbox{ (stat.)} \pm 0.03\mbox{ (syst.)}$
and $\delta_{\perp} =  2.97 \pm 0.06 \mbox{ (stat.)} \pm 0.01\mbox{ (syst.)}$, 
which are in agreement and competitive with the current B factories results
\cite{Aubert:2004cp}.

We present the first Tevatron studies of the $B_s^0 \to J/\psi \phi$ decay mode 
when the initial state of the $B_s^0$ meson is identified exploiting the flavor 
tagging information. In fact, such information allows to separate the time 
evolution of mesons originally produced as $B_s^0$ or $\bar{B}_s^0$. The angular
analyses which do not use the flavor tagging are sensitive to $|\cos(2\beta_s)|$
and $|\sin(2\beta_s)|$, leading to a $4$-fold ambiguity in the likelihood for 
the determination of $2\beta_s$ (see Figure~\ref{fig:untagged}). 
On the other hand, utilizing flavor tagging algorithms, the analyses gain 
sensitivity to the sign of $\sin(2\beta_s)$ reducing by half the allowed region
for $\beta_s$. 
CDF performed a flavor tagged analysis on a $1.35~\mathrm{fb}^{-1}$ data-set of 
$B_s^0 \to J/\psi \phi$ reconstructed events, which yields $\simeq 2,000$ 
signal candidates\cite{Aaltonen:2007he}. 
The measured efficiencies for OST and SSKT are $\epsilon_{OST}=(96\pm1)\%$ and 
$\epsilon_{OST}=(50\pm1)\%$. The dilutions are respectively 
$\mathcal{D}_{OST}=(11 \pm 2)\%$ for the OST and $\mathcal{D}_{SSKT}=(27\pm4)\%$
for the SSKT. 
The addition of tagging information improves the regularity 
of the likelihood with respect to the untagged case, but still non-Gaussian 
uncertainties and biases are observed in simulated experiments with the 
available statistics. Thus, CDF reports a confidence region constructed 
according to the Feldman Cousins criterion with rigorous inclusion of 
systematics uncertainties. 
In fact, any $\Delta\Gamma_s-\beta_s$ pair is excluded at a given
CL only if it can be excluded for any choice of \emph{all} other fit parameters,
sampled uniformly within $\pm 5~\sigma$ of the values determined in their 
estimate on data. Assuming the Standard Model predicted values of 
$2\beta_s=0.04~\mathrm{rad}$ and $\Delta \Gamma_s = 0.096~\mathrm{ps}^{-1}$, the
probability of a deviation as large as the observed data is $15\%$, which 
corresponds to $1.5$ Gaussian standard deviations. 
Moreover, if $\Delta\Gamma_s$ is treated as a nuisance parameter, 
thus fitting only for $2\beta_s$, CDF finds $2\beta_s \in [0.31,2.82]
~\mathrm{rad}$ at the $68\%$ confidence level.
By exploiting the current experimental and theoretical information, CDF extracts
tighter bounds on the {\sf CP} violation phase $\beta_s$. Imposing the 
constraint on $|\Gamma_{12}|=0.048 \pm 0.018~\mathrm{ps}^{-1}$
in $\Delta\Gamma_s=2|\Gamma_{12}|\cos(2\beta_s)$~\cite{Lenz:2006hd},
$2\beta_s \in [0.24,~1.36] \cup [1.78,~2.90]~\mathrm{rad}$ at the 
$68\%$ CL. Additionally constraining the strong phases $\delta_{\parallel}$ and 
$\delta_{\perp}$ to the $B$ factories results on 
$B^0 \to J/\psi K^{*0}$~\cite{Aubert:2004cp} and the $B_s^0$ mean width to the 
world average $B^0$ width~\cite{Yao:2006px}, it is found 
$2\beta_s \in [0.40,~1.20]~\mathrm{rad}$ at the $68\%$ CL.
The D\O~Collaboration reports an analysis~\cite{2008fj} on $2,000$ signal 
$B_s^0 \to J/\psi \phi$ candidates, reconstructed in $2.8~\mathrm{fb}^{-1}$. 
D\O~combines the tagging algorithms, as done in their $B_s^0$ mixing analysis.  
The total tagging power is $\epsilon \mathcal{D}^2 =(4.68 \pm 0.54)\%$ and a tag
is defined for $99.7\%$ of the events.
To overcome the likelihood pathologies described above, D\O~decides to vary the
strong phases around the world-averaged values for the $B^0 \to J/\psi K^{*0}$ 
decay~\cite{Barberio:2007cr}, applying a Gaussian constraint. 
This removes the $2$-fold ambiguity, inherent the measurement for arbitrary 
strong phases. The strong phases in $B^0 \to J/\psi K^{*0}$ and 
$B_s^0 \to J/\psi \phi$ cannot be exactly related in the $SU(3)$ limit, so the 
width of the Gaussian is chosen to be $\pi/5$, allowing for some degree of 
$SU(3)$ symmetry violation. The fit with all floating parameters yields to the
measurements

\bea
\phi_s & = & -0.57 ^{+0.24}_{-0.30}        \mbox{ (stat.)} 
                   ^{+0.07}_{-0.02}        \mbox{  (syst.)}~\mathrm{rad},
                                                           \nonumber \\ 
\Delta\Gamma_s & = & 0.19 \pm 0.07         \mbox{ (stat.)} 
                          ^{+0.02}_{-0.01} \mbox{  (syst.)}~\mathrm{ps}^{-1}, 
                                                           \nonumber \\
\tau_s & = & 1.52 \pm 0.05 \mbox{ (stat.)} \pm 0.01 \mbox{ (syst.)}
                                                           ~\mathrm{ps}.
\eea

\begin{figure}
\begin{center}
\epsfig{figure=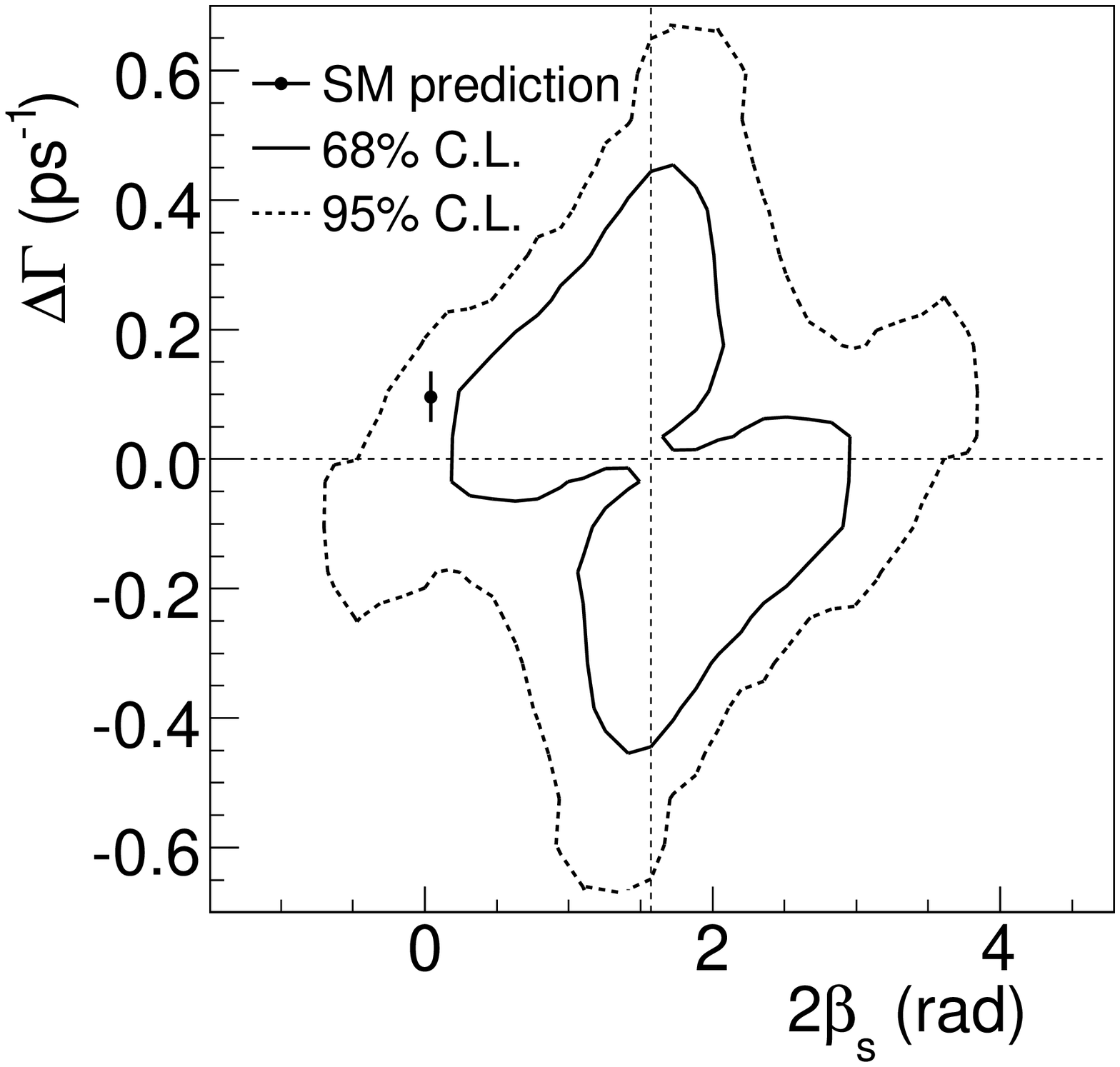, height=1.9in, width=2.2in}
\epsfig{figure=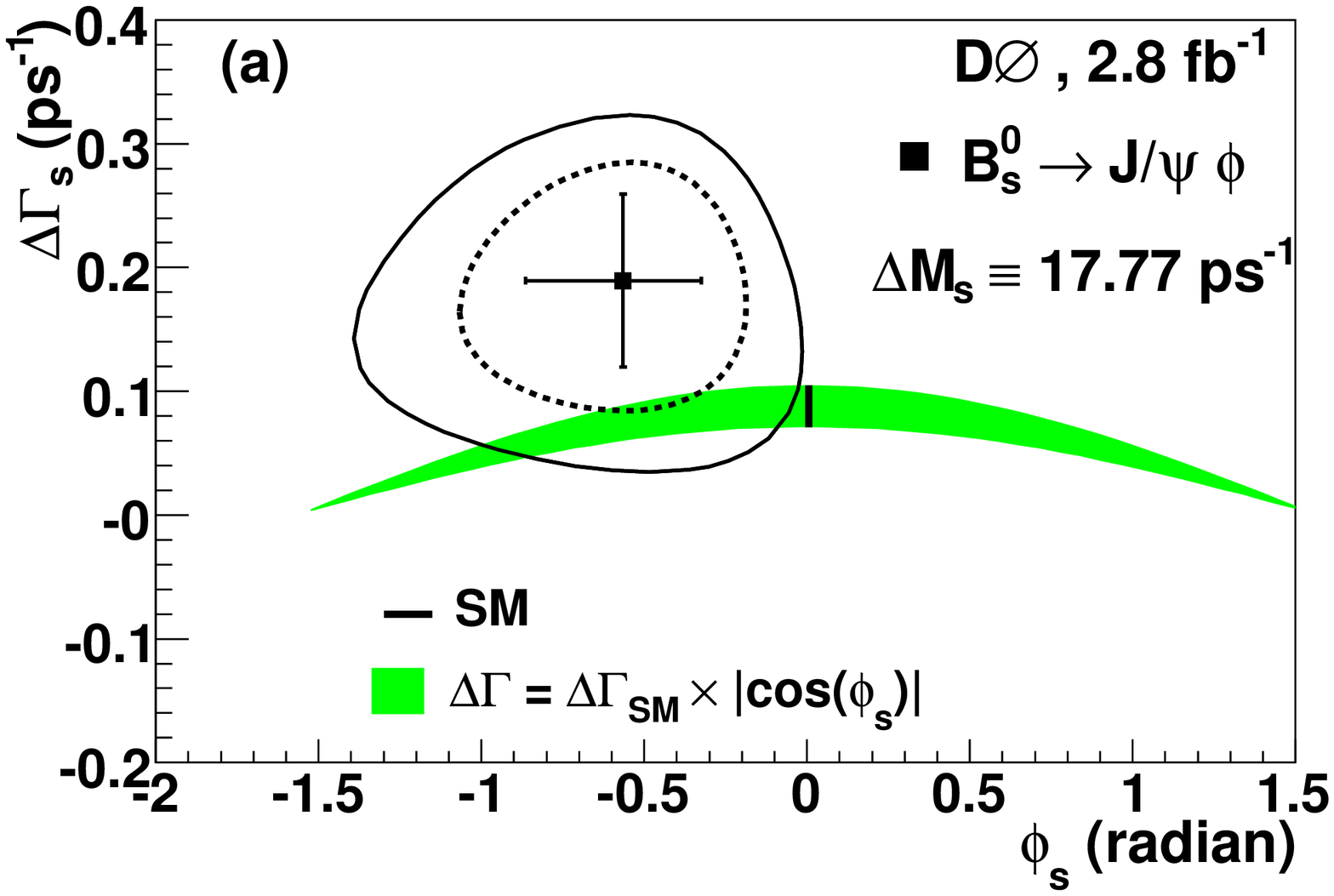               , height=1.8in, width=2.2in}
\caption{Left: CDF confidence level contour in the plane 
  $2\beta_s-\Delta\Gamma_s$ when using flavor tagging. It quotes the $68\%$ 
  (solid) and $95\%$ (dashed) C.L. The solution with 
  $\Delta \Gamma_s > 0$ corresponds to $\cos(\delta_{\perp}) < 0$ and 
  $\cos(\delta_{\perp} - \delta_{\parallel}) > 0$. The opposite is true for the 
  solution with $\Delta \Gamma_s < 0$. Right: D\O~confidence level contours in 
  the $\Delta\Gamma_s-\phi_s$ plane. The curves correspond to the expected 
  $68\%$ (dashed) and $90\%$ (solid) CL. The cross represents the best 
  estimate fit with the one-dimensional uncertainties. According to D\O~
  nomenclature, $\phi_s=-2\beta_s$. \hspace{4.6cm}
  \label{fig:tagged}}
\end{center}
\end{figure}

The allowed ranges at the $90\%$ CL for the parameters of interest are found to 
be $\phi_s \in [-1.20, 0.06]~\mathrm{rad}$ and 
$\Delta\Gamma_s \in [0.06, 0.30]~\mathrm{ps}^{-1}$. 
The expected confidence level contours in the $\phi_s-\beta_s$ plane at $68\%$ 
and $90\%$ CL are depicted in Figure~\ref{fig:tagged}. The level of agreement 
with the Standard Model corresponds to $6.6\%$, which is obtained by generating 
pseudo-experiments with the initial value for $\phi_s$ set to 
$-0.04~\mathrm{rad}$ and counting the events whose obtained fitted value of the 
phase is lower than the measured $-0.57~\mathrm{rad}$. The results supersede the
previous D\O~untagged analysis on a smaller $B_s^0 \to J/\psi \phi$ sample.

%===============================================================================
\section{Charge Asymmetry in $B_s^0$ Semileptonic Decays}
\label{sec:bsasymm}
%===============================================================================
Another way of studying the {\sf CP} violation induced by the $B_s$ mixing, 
is to measure the charge asymmetry in semileptonically decaying mesons. The 
charge asymmetry is connected to the {\sf CP} violating phase $\phi_s$, through 
the relationship $A^s_{SL}=\Delta\Gamma_s/\Delta m_s \times tan(\phi_s)$.
With the underlying assumption of $\phi_s=-2\beta_s$ (see Section
~\ref{sec:bspheno}), an independent measurements on charge asymmetry could be 
used to constrain the {\sf CP} violating phase $\beta_s$~\cite{Abazov:2007zj}.
D\O~Collaboration performed two independent analyses to extract $A^s_{SL}$. The 
first result is based on the di-muon charge asymmetry measurement
~\cite{Abazov:2006qw}, defined as
\be
A^{\mu \mu}_{SL}= \frac{N(bb \to \mu^+ \mu^+) - N(bb \to \mu^- \mu^-)}
                       {N(bb \to \mu^+ \mu^+) + N(bb \to \mu^- \mu^-)}.
\ee
The following asymmetry gets its contributions from both $B^0$ and $B_s^0$: by 
using the world average value for $B^0$ and $B_s^0$ production fractions and the
$B^0$ charge asymmetry measurements from the $B$ factories, D\O~extracts the 
$B_s^0$ charge asymmetry on a data-set of $1.0~\mathrm{fb}^{-1}$:

\be
A^{\mu \mu, B_s^0}_{SL} = -0.0064 \pm 0.0101 \mbox{ (stat. $+$ syst.)}.
\ee

CDF~Collaboration also released a similar measurement of the di-muon charge 
asymmetry~\cite{asymmCDF} on a sample of $1.6~\mathrm{fb}^{-1}$ data. In this 
analysis, the unbinned likelihood is performed using the impact parameter 
information of the two muons, in order to separate the $b-\bar{b}$ component of 
the sample from the others which arise from prompt and charm sources:

\be
A^{\mu \mu, B_s^0}_{SL} = 0.020 \pm 0.021 \mbox{  (stat.)  } 
                                \pm 0.016 \mbox{  (syst.)  }
                                \pm 0.009 \mbox{ (inputs) }.
\ee

Additionally to the statistical and systematic uncertainties, the last
uncertainty term comes from the world average value for $B^0$ and $B_s^0$ 
production fractions and the $B^0$ charge asymmetry measurements already 
discussed in the description of D\O~results. Compared to CDF, D\O~analysis has 
strongly reduced systematics uncertainties thanks to a regular flipping of the 
magnet polarity. Such technique, removing most of the artificial asymmetry in 
the detector response, is constantly used by D\O~to measure all the charge 
asymmetries described along this paper.

The D\O~Collaboration probes the $\phi_s$ phase also by measuring the charge 
asymmetry in an untagged sample of $B_s^0 \to \mu D_s$ decays, with 
$D_s \to \phi(\to K^+ K^-) \pi$. With a data-set of 
$1.3~\mathrm{fb}^{-1}$ the charge asymmetry is found to be~\cite{Abazov:2007nw}

\be
A^{\mu D_s}_{SL} = 0.0245 \pm 0.0193 \mbox{ (stat.)} 
                          \pm 0.0035 \mbox{ (syst.)}.
\ee

%===============================================================================
\section{Charge Asymmetry in $B^+ \to J/\psi K^+$ Decay}
\label{sec:buasymm}
%===============================================================================
We present a search for direct {\sf CP} violation in $B^+ \to J/\psi K^+$ decays
~\cite{Abazov:2008gs}. The event sample is selected from $2.8~\mathrm{fb}^{-1}$ 
of $p\bar{p}$ collisions recorded by D\O~experiment. The charge asymmetry is 
defined as

\be
A_{CP}(B^+ \to J/\psi K^+)=\frac{N(B^- \to J/\psi K^-) - N(B^+ \to J/\psi K^+)}
                                {N(B^- \to J/\psi K^-) + N(B^+ \to J/\psi K^+)}.
\ee

By using a sample of approximately $40,000~B^+ \to J/\psi K^+$ decays, the 
asymmetry is measured to be 
$A_{CP}=0.0075 \pm 0.0061 \mbox{ (stat.)} \pm 0.0027 \mbox{ (syst.)}$. 
The result is consistent with the world average~\cite{Yao:2006px} and the 
Standard Model expectation $A_{CP}(B^+ \to J/\psi K^+) \simeq 0.003$
~\cite{Hou:2006du}, but has a factor of two improvement in precision, thus 
representing the most stringent bound for new models which predict large values
of $A_{CP}(B^+ \to J/\psi K^+)$. Furthermore, D\O~provides the direct {\sf CP} 
violating asymmetry measurement in $B^+ \to J/\psi \pi^+$, 
$A_{CP}(B^+ \to J/\psi\pi^+)= -0.09 \pm 0.08 \mbox{ (stat.)} \pm 0.03
\mbox{ (syst.)}$. The result agrees with the previous measurements of this 
asymmetry~\cite{Yao:2006px} and has a competitive precision.

%===============================================================================
\section{Conclusions}
\label{sec:concl}
%===============================================================================
After the successful $B_s^0$ oscillation observation, the CDF and D\O~
Collaboration directed 
their effort in the exploration of the mixing-induced {\sf CP} violation effect 
in the $B_s^0$ system. 
We described the first tagged measurement in $B_s^0 \to J/\psi \phi$ performed
at the CDF II detector, which improved the sensitivity to the {\sf CP} violating
phase $\beta_s$, excluding negative and large values for the phase itself. The
D\O~Collaboration promptly delivered a similar analysis confirming the results.
The agreement of the analyses of $B_s^0 \to J/\psi \phi$ decays, 
shows an interesting fluctuations in the same direction from CDF and D\O~
experiments and they will certainly need further investigations to support an 
evidence, which would be possible exploiting the full Run II data sample, if 
these first indications are confirmed in the future.
We also reviewed the charge asymmetry measurements of $B_s^0$ semileptonic 
decays, which provide another independent test for the {\sf CP} violation in 
$B_s^0$ mixing and can be combined with the analyses on $B_s^0 \to J/\psi \phi$ 
to get a better understanding of the {\sf CP} violating phenomena. 
Finally, we presented the world most precise direct {\sf CP} violating asymmetry
in the $B^+ \to J/\psi K^+$ decay mode.
The Tevatron experiments are becoming increasingly competitive with B factories
results on $B^0/B^+$ decays and complementary to them in corresponding $B_s^0$
modes. Since many of the analyses reported do not even 
use half of the statistics available, significant improvements are
expected in the future, as the Tevatron keeps producing data.

\end{document}